\documentclass[sigconf]{acmart}

\usepackage{CJK}
\usepackage{graphicx}
\usepackage{caption}
\usepackage{epstopdf}
\usepackage{booktabs} 
\usepackage{subfigure}

\setcopyright{acmcopyright}

\acmConference[InTI'2018]{SIGIR Workshop on Intelligent Transportation Informatics}{July 12, 2018}{Ann Arbor, Michigan, USA}
\acmYear{2018}
\copyrightyear{2018}

\begin{document}
\title{POI Semantic Model with a Deep Convolutional Structure}

\author{Ji Zhao, Meiyu Yu$^{*}$, Huan Chen, Boning Li, Lingyu Zhang, Qi Song, Li Ma, Hua Chai, Jieping Ye}
\thanks{$^{*}$Meiyu Yu participated in this work when she worked as an intern in DiDi Chuxing}
\affiliation{
	\institution{DiDi Chuxing, Beijing, China}
}
\email{{zhaojijet, yumeiyu_i, chenhuan, liboning, zhanglingyu, songqihx, malimarey, chaihua, yejieping} @didichuxing.com}

\renewcommand{\shortauthors}{Ji Zhao et al.}

\begin{abstract}
When using the electronic map, POI\footnote[1]{POI: A point of interest, is a specific point location that someone may find useful or interesting.} retrieval is the initial and important step, whose quality directly affects the user experience. Similarity between user query and POI information is the most critical feature in POI retrieval. An accurate similarity calculation is challenging since the mismatch between a query and a retrieval text may exist in the case of a mistyped query or an alias inquiry. In this paper, we propose a POI latent semantic model based on deep networks, which can effectively extract query features and POI information features for the similarity calculation. Our model describes the semantic information of complex texts at multiple layers, and achieves multi-field matches by modeling POI's name and detailed address respectively. Our model is evaluated by the POI retrieval ranking datasets, including the labeled data of relevance and real-world user click data in POI retrieval. Results show that our model significantly outperforms our competitors in POI retrieval ranking tasks. The proposed algorithm has become a critical component of an online system serving millions of people everyday.
\end{abstract}

%
%
%

\keywords{POI, Convolution, Neural Network, Semantic Model}

\maketitle

\section{Introduction}

\par
The electronic map, is now indispensable in our daily commute and the POI retrieval system is one of its critical components. When user input a query, the system will retrieve and return POI results according to relevance between the query and the POI information. Thus the main problem in information retrieval system is text matching. Compared with the lexical matching, the semantic matching should be paid more attention in text matching of the retrieval system, because the same concept may be expressed in different words or styles in the query and the retrieval text. 

\par
Recently, semantic models based on neural networks have been proposed for information retrieval, such as DSSM~\cite{huang2013learning}, CLSM~\cite{shen2014latent}, etc. Although the models mentioned above have been successfully applied in relevance retrieval, these methods cannot fully satisfy our demands. In spite of incompletion of query, users can acquire instant results while typing in the POI retrieval system. Due to the characteristics of the Chinese Pinyin\footnote[2]{Pinyin: Hanyu Pinyin Romanization, often abbreviated to Pinyin, is the official romanization system for Standard Chinese in China.} IME (input method editor), the input query includes Chinese characters and Pinyin; such complicated query combinations resulting from incomplete search can appear in the POI retrieval. In addition, the POI text information consists of the name and detailed address, which will be fully used by our retrieval system. To address all of these challenges, we propose DPSM (Deep POI Semantic Model) in this paper, shown in Figure~\ref{fig:query_poi_model}. DPSM is mainly composed of two models including the Query Model and the POI Model. When inputing a query by user, we retrieve POI from our POI database, DPSM projects the query to vector $\mathbf{Q}$ and projects the POI including the POI name and POI address to vector $\mathbf{P}$. Finally, DPSM outputs the cosine similarity of $\mathbf{Q}$ and $\mathbf{P}$. The DPSM we propose performs well on an offline test set. DPSM also has outstanding performance in runtime and has been deployed to a production system serving millions of people everyday.

\par
The properties of our model are as follows:
\begin{itemize}
	\item We use letter and word granularity to present the text for capturing incomplete query information and extracting more meaningful semantic information.
	\item We insert a word embedding layer before the convolution layer to decrease the convolution kernel size.
	\item We model the POI name and POI address respectively to obtain multi-field information of POI and improving the relevance of POI retrieval.
\end{itemize}

\section{Deep POI Semantic Model}
\label{sec:deep_poi_semantic_model}

\par
\begin{figure}[!tbp]
	\begin{center}
		\includegraphics[width=0.9\linewidth]{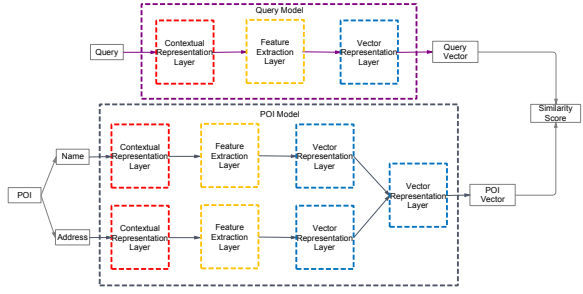}
	\end{center}
	\caption{The proposed Query Model (purple box) and POI Model (dark gray box) in DPSM.}
\label{fig:query_poi_model}
\end{figure}

In this section, we will introduce our model, shown in Figure~\ref{fig:query_poi_model}. We first show how the Query Model uses the feature vector to represent the query information, then we show how the POI Model represents the POI information as a feature vector, finally we introduce the loss function of our Deep POI Semantic Model.

\subsection{Query Model}
\label{sec:query_model}
\par
\begin{figure}[!tbp]
	\begin{center}
		\includegraphics[width=0.9\linewidth]{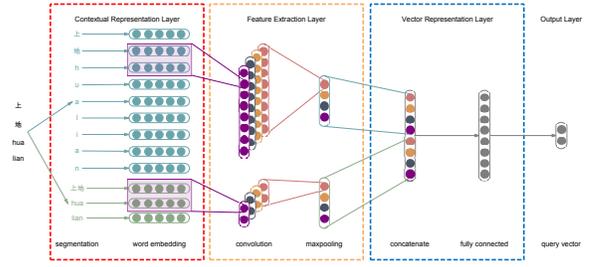}
	\end{center}
	\caption{The proposed Query Model. The block included in the red box is the Contextual Representation Layer, including query information expressed with the letter granularity (in turkish blue) and the word granularity (in foliage green) and projected to vectors. The block included in the yellow box is the Feature Extraction Layer, including n-gram features extracted in the convolution layer and the maxpooling layer. The block included in the blue box is the Vector Representation Layer, including the high-level features extracted in the fully connected layer.}
	\label{fig:query_model_1}
\end{figure}

The network structure of the Query Model as shown in Figure~\ref{fig:query_model_1} includes (1) a Contextual Representation Layer; (2) a Feature Extraction Layer; (3) a Vector Representation Layer. 

\subsubsection*{Contextual Representation Layer}
\label{sec:contextural_representation_layer}
\par
This layer includes multi-granularity representation and embedding.
\par
There are many incomplete queries or queries mixed with Pinyin inputs by users in our POI retrieval system. Therefore, we use multi-granularity representation for the query, including the letter granularity representation and the word granularity representation. The letter granularity representation can describe better than word granularity representation in incomplete or mixed with Pinyin inputs queries since the word granularity representation is likely to be out-of-vocabulary in those queries. In the Chinese context, the contextual information of words is usually more than that of letters and a simple combination of letter features cannot represent the features of corresponding words well. We use the Hidden Markov Model~\cite{zhang2003hhmm} to segment query to get word granularity representation.

\par
After obtaining query representation of letter granularity and word granularity, we use embedding matrix to project query information to a feature vector. In Chinese setting, if we use the letter-tri-gram method like CLSM, we generate a large size of vocabulary. It is too large for computing. So, we apply the embedding method to project letters or words to a low-dimensional vector. Based on the name and address of POI data, we count the high-frequency letters and words, and generate the embedding matrix. Through the embedding matrix, we can project query information expressed in letter granularity and word granularity.

\subsubsection*{Feature Extraction Layer}
\label{sec:feature_extraction_layer}

\par
This Layer includes a convolution operation and a maxpooling operation.

\par
The convolution operation on the embedding vector can be seen as the extraction of n-gram features of the query through a sliding window. In order to obtain all embedding information in a sliding window, we use a convolution kernel $f$ as wide as the embedding matrix and extract the corresponding n-gram features by changing the length of the convolution kernel. Finally, we use Relu activation function to get nonlinear n-gram features $\mathbf{c}=<\mathbf{c_l},\mathbf{c_w}>$. The whole process is shown in Eq.~\ref{convolution_operation} and Eq.~\ref{Relu_activation}, where  $\otimes$ is the convolution operator, $i = \mathbf{l}, \mathbf{w}$ is represented as the letter vector and the word vector, and $\mathbf{b_c}$ is the bias vector. In the experiment, we try tanh and Relu activation functions. Our experimental results show that the Relu activation function performs better than the tanh activation function for our model, because the Relu activation function makes the network converge faster.

\begin{equation}
\label{convolution_operation}
\mathbf{c_i} = \mathrm{Relu} \left(\mathbf{q_i} \otimes {f}+\mathbf{b_c} \right),  i = \mathbf{l},\mathbf{w}
\end{equation}
\begin{equation}
\label{Relu_activation}
\mathrm{Relu}\left(\mathbf{x}\right) = max\left(0, \mathbf{x}\right)
\end{equation}

Because the length of query varies, the length of the convolution vector output from the convolution layer also varies. We use the pooling method to unify the length of convolution vectors. Another advantage of pooling is that pooling can extract contextual information from the whole text. In order to highlight the difference of text semantic features, we use maxpooling in the experiment to take the maximum value of features in the same convolution channel to get maxpooling features $\mathbf{pool}=<\mathbf{pool_l},\mathbf{pool_w}>$, shown in Eq.~\ref{pooling_features}:

\begin{equation}
\label{pooling_features}
\mathbf{pool_i}=\max{\left(\mathbf{c_i}\right),i=\mathbf{l},\mathbf{w}}
\end{equation}

\subsubsection*{Vector Representation Layer}
\label{sec:vector_representation_layer}

\par
This Layer includes a fully connected layer and an output layer.

\par
We add a fully connected layer after the maxpooling layer to further extract high-dimensional features as well as an output layer formed with Sigmoid activation function to project the final semantic feature vector, as shown in Eq.~\ref{high_level_features_1} and Eq.~\ref{high_level_features_2}:

\begin{equation}
\label{high_level_features_1}
\mathbf{h} = \mathrm{Relu}\left( \mathbf{w_h}<\mathbf{pool_l};\mathbf{pool_w}> + \mathbf{b_h}\right)
\end{equation}
\begin{equation}
\label{high_level_features_2}
\mathbf{L}=\mathrm{Sigmoid} \left( \mathbf{w_L}\mathbf{h} + \mathbf{b_L} \right)
\end{equation}

In Eq.~\ref{high_level_features_1}, $\mathbf{h}$ is the fully connected layer; $\mathbf{w_h}$  is the weight of the fully connected layer; and $\mathbf{b_h}$ is the bias of the fully connected layer. The activation function of the output layer $\mathbf{L}$ is Sigmoid function. The fully connected layer can model the relationship of letter granularity and word granularity, which can unify the multi-granularity features and extract the high-dimensional features of query.

\subsection{POI Model}
\label{sec:poi_model}

\par
In the POI retrieval scenario, the query input by user is either the POI name or the POI address. So for POI retrieval system, we use both POI name and POI address to build our POI Model, shown in Figure~\ref{fig:query_poi_model}.

\par
POI Model also has three layers. The Contextual Representation Layer and the Feature Extraction Layer are the same as the Query Model; in the Feature Extraction Layer, the convolution kernel of POI Model shares the same weights with the convolution kernel in the Query Model;  in the Vector Representation Layer, we first concatenate the pooling features of POI name and POI address then use a fully connected operation to extract high-dimensional features representing POI information. In order to keep in line with the dimensions of the query vector, we project it to 128 dimensions.

\par
For a given query and a given POI, we use the Query Model and the POI Model to get corresponding feature vector respectively and calculate the cosine similarity of the two feature vectors to get a relevance score.

\subsection{Loss Function}
\label{sec:loss_function}

\par
In the process of training Deep POI Semantic Model, we use the softmax loss function which is widely used in Learn To Rank model~\cite{burges2005learning}. The softmax loss function is shown in Eq.~\ref{positive_negative} and Eq.~\ref{loss_function}:

\begin{equation}
\footnotesize
\label{positive_negative}
P\left({\mathbf{P}^+} \mid \mathbf{Q} \right) = \frac {exp{\left(\gamma sim\left(\mathbf{Q},{\mathbf{P}^+}\right)\right)}} {\sum_{\mathbf{P}^{'} \in \mathbf{P}}{exp{\left(\gamma sim\left(\mathbf{Q},{\mathbf{P}^{'}}\right)\right)}} }
\end{equation}
\begin{equation}
\footnotesize
\label{loss_function}
Loss = -log{\prod\limits_{\left(\mathbf{Q},{\mathbf{P}^+}\right)}} P\left(\mathbf{P}^+ \mid \mathbf{Q}\right)
\end{equation}

Where $\mathbf{Q}$ is the input query; $\mathbf{P}^+$ is the positive sample; $\mathbf{P}^-$ is the negative sample and is randomly generated by noise contrastive estimation (NCE)~\cite{gutmann2012noise} method according to the popularity of POI; $\mathbf{P}$  is the corresponding sample set of the query retrieval; sim function is the cosine similarity function; $\gamma$ is the smooth coefficient.

\section{Experiments}
\label{sec:experiments}

\subsection{Datasets}
\label{sec:dataset}

\par
We first introduce our training dataset and testing dataset. In the training dataset, we randomly select 10 million real clicked POIs from our app's one-month logs as positive samples. When users choose POIs, they are inclined to click POIs which are on the top of the displayed list; this phenomenon is called ``location bias". So we use the NCE method to randomly generate negative samples based on the POI popularity. There are two types of test datasets. One is the click data; we collect 1 million POI clicks from logs and this kind of data can test our model in the location bias scenario. Another one is the relevance data; we annotate 60,000 queries and POIs. Such kind of data can test our model without selection bias. In order to better understand our experimental data collected by the POI retrieval system, we calculate the average length of query, POI name and POI address in the training dataset and the testing dataset and the percentage of queries with Pinyin in the click dataset as well. The statistics are shown in Table~\ref{tab:statistic_distribution_of_datasets}. As shown in Table~\ref{tab:statistic_distribution_of_datasets}, the percentage of queries with Pinyin in the click dataset is 13.42\% and in the relevance dataset is 9.67\%.  Moreover, the average length of query is 3.92 in the click dataset and 3.25 in the relevance dataset. We can observe that the majority of queries are short queries and about 10\% of them are mixed with Pinyin.

\begin{table}[!tbp]
	\newcommand{\tabincell}[2]{\begin{tabular}{@{}#1@{}}#2\end{tabular}}
	\small
	\centering
	\caption{Distributions of Datasets.}
	\label{tab:statistic_distribution_of_datasets}
	\begin{tabular}{c|c|c}
		\toprule
		\tabincell{c}{Query Percentage} & \tabincell{c}{Click Dataset} & \tabincell{c}{Relevance Dataset}\\
		\midrule
		\tabincell{c}{with Pinyin}& 13.42\% & 9.67\%\\
		\tabincell{c}{without Pinyin} & 86.58\% & 90.37\%\\
		\bottomrule
		\toprule
		\tabincell{c}{Average Length} & \tabincell{c}{Click Dataset} & \tabincell{c}{Relevance Dataset}\\
		\midrule
		\tabincell{c}{Query}& 3.92 & 3.25\\
		\tabincell{c}{POI name} & 8.26 & 8.69\\
		\tabincell{c}{POI address} & 19.01 & 19.04\\
		\bottomrule
	\end{tabular}
\end{table}

\subsection{Baseline and Experiment Setup}
\label{sec:baseline_and_experiment_settings}
\par
We compare five deep learning models in the datasets mentioned above. We choose NDCG score to evaluate model performance.

\par
\textbf{DSSM}: DSSM is a state-of-the-art deep learning model for web retrieval. We first count the high-frequency Chinese characters and Pinyin alphabets to constitute the vocabulary, resulting in a total of 10,000 characters. Then we generate the one-hot feature based on the query and POI name information by a word hashing layer with the vocabulary. Finally, we obtain the feature vector of the query and the POI name through a fully connected layer and calculate their cosine similarity.

\par
\textbf{CDESM}: CLSM is similar to DSSM for web retrieval. Compared with DSSM, CLSM has the convolution operator to extract contextual features.
To solve the problem, we first use 10,000 high-frequency words to form a vocabulary and we use an embedding layer based on this vocabulary to project texts to feature vectors. Then we extract n-gram features from the convolution layer, and finally get the feature vector of the query and POI name by a fully connected layer. We call the model CDESM.

\par
\textbf{CDESM+WS}: In order to get more semantic information, we represent the text information by using both letter granularity and word granularity, and enlarge the vocabulary to 300,000 by adding diverse granularity information. We call the CDESM which involves the letter granularity and word granularity as CDESM+WS (CDESM+word segmentation).

\par
\textbf{CDESM+WS+ADDR}: We involve POI address information on the basis of CDESM+WS. In the model, we extend the POI name with the POI address directly and we take the same operation with CDESM+WS to extract POI features.

\par
\textbf{DPSM}: DPSM is the model proposed in this paper.
The size of trainable embedding matrix is 300,000*100; the size of the convolution kernel is 2*100*300; the convolution stride is 1, and one fully connected layer is added after the maxpooling layer.

\subsection{Results}
\label{sec:results}
\par
We next compare five competing deep learning models. 

\par
Table~\ref{tab:offline_click_experimental_results} shows the experimental results on the click dataset. We adopt DSSM as our baseline. 
In CDESM, it uses the embedding and convolution operation to extract contextual information, so the curve of CDESM's loss is more smooth and the NDCG score of CDESM is a little better than DSSM. CDESM+WS has a better performance than the two preceding models. And CDESM+WS+ADDR achieves better performance than that of CDESM+WS; this is due to the use of the POI address to model input which can help make a better use of the POI. Compared with the previous four models, we can observe from Table~\ref{tab:offline_click_experimental_results} that our proposed model, DPSM, achieves the best performance. This is probably due to (1) DPSM uses multi-granularity features of texts, and (2) it exploits features of the POI address more effectively.

\begin{table}[!tbp]
	\centering
	\small
	\caption{Offline Click Experimental Results. Comparative experimental results of DPSM with DSSM, CDESM, CDESM+WS and CDESM+WS+ADDR.}
	\label{tab:offline_click_experimental_results}
	\begin{tabular}{c|c|c}
		\toprule
		Models & NDCG@3 & NDCG@10\\
		\midrule
		DSSM & 0.7319 & 0.8726\\
		CDESM & 0.7355 & 0.8743\\
		CDESM+WS & 0.7435 & 0.8776\\
		CDESM+WS+ADDR & 0.7444 & 0.8776\\
		DPSM & 0.7524 & 0.8812\\
		\bottomrule
	\end{tabular}
\end{table}

\begin{table}[!tbp]
	\centering
	\small
	\caption{Offline Relevance Experimental Results. Comparative experimental results of DPSM with DSSM, CDESM, CDESM+WS and CDESM+WS+ADDR.}
	\label{tab:offline_relevance_experimental_results}
	\begin{tabular}{c|c|c}
		\toprule
		Models & NDCG@3 & NDCG@10\\
		\midrule
		DSSM & 0.6021 & 0.6099\\
		CDESM & 0.6080 & 0.6188\\
		CDESM+WS & 0.6198 & 0.6256\\
		CDESM+WS+ADDR & 0.6203 & 0.6212\\
		DPSM & 0.6231 & 0.6333\\
		\bottomrule
	\end{tabular}
\end{table}

\par
Click data represents users' choices in POI retrieval, but it is influenced by the ranking order of the POI list and users' personal preferences. We include an additional experiment using a relevance dataset labeled by humans and the experimental results are showed in Table~\ref{tab:offline_relevance_experimental_results}. The results of the relevance case are similar to click case. DPSM performs the best in the experiments.
\par
In addition, DPSM is applied to GBDT~\cite{friedman2001greedy}(online ranker) as a feature through Tensorflow Serving~\cite{abadi2016tensorflow} and the relevance feature is ranked as the most effective feature. 
In the test stage, we give Tensorflow-Serving a burden request of 1,000 queries per second(qps), and the response time of DPSM is about 15ms, which can be used directly for online services. We also use 60,000 relevance data to test online models with DPSM and gain the improvement of 3\% in terms of NDCG@10.

\subsection{Analysis}
\label{sec:analysis}
\par
\begin{figure}[!tbp]
	\begin{center}
		\includegraphics[height=0.2\textheight]{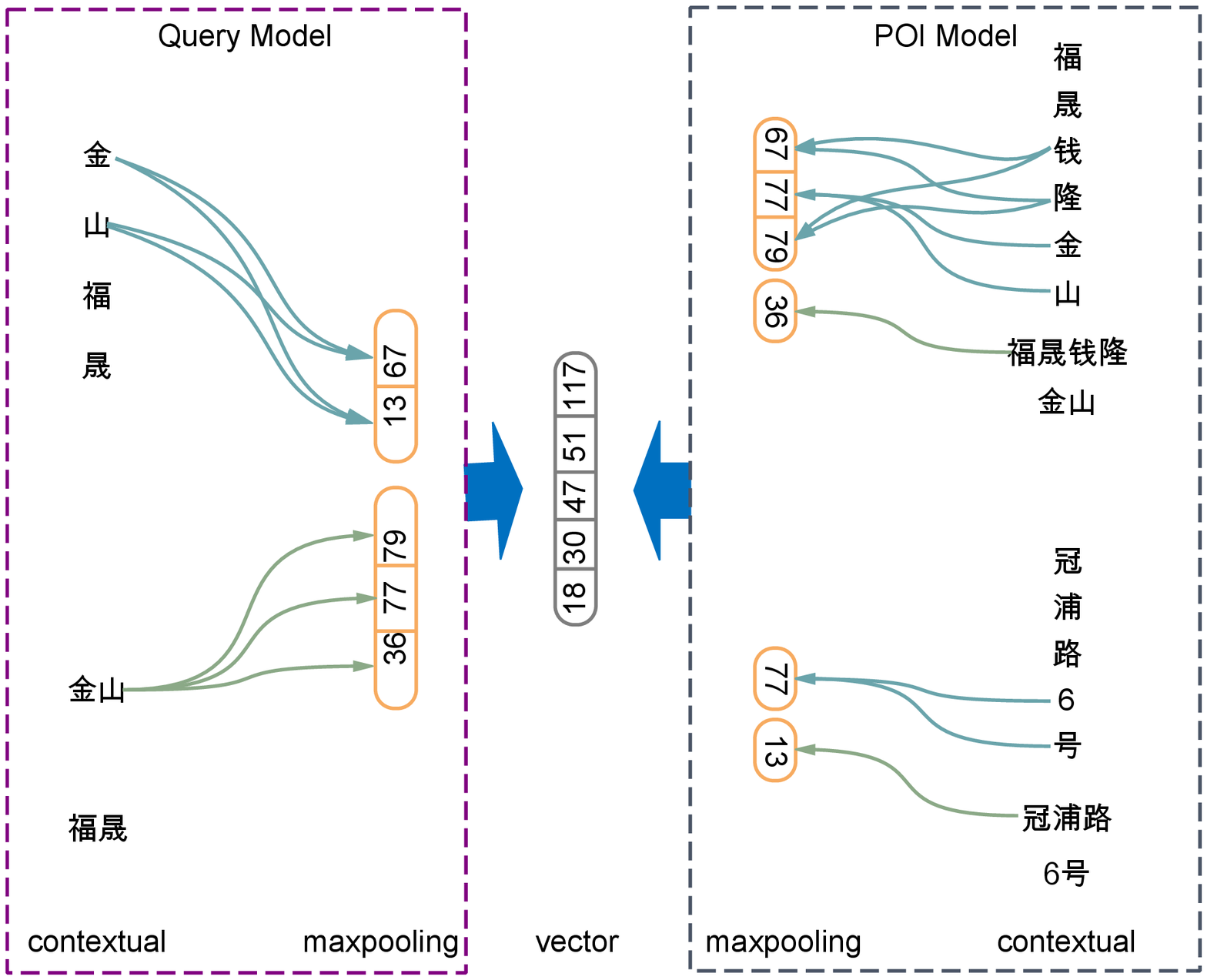}
	\end{center}
	\caption{DPSM Data Flow. The purple box is the Query Model and the dark gray box is the POI Model. The contextual layer shows that DPSM extracts contextual features of the query, POI name and POI address; the maxpooling layer represents the indices of high activation neurons in maxpooling layer which is produced by convolution and maxpooling on contextual feature; the vector layer shows the indices of high activation neurons in query vector and POI vector.}
	\label{fig:case_explanation_1}
\end{figure}

In order to better understand how DPSM works, we analyze the data flow of DPSM. Figure~\ref{fig:case_explanation_1} shows the detailed information of DPSM. The purple box represents the Query Model and the dark gray box represents the POI Model. We show three important layers which can explain how the similarity score  between query and POI is computed: contextual layer, maxpooling layer and vector layer. In contextual layer, DPSM extracts contextual features by convolution. For better analysis, we trace the activation of neurons at maxpooling layer as semantic information and analyze the relationship between the query and the POI semantic information. We find that indices of high activation neurons in the maxpooling layer between the query and POI name is 36, 67, 77, 79 and common indices of high activation neurons in the maxpooling layer between the query and the POI address is 13, 77. In our DPSM, the layer between maxpooling and the feature representation vector is a fully connected layer, so the activation neuron contributes to the feature significance in the feature representation vector. We also trace the activation neurons in the feature representation vector. The common indices of high activation neurons in the query maxpooling layer and POI maxpooling layer contribute to the common indices of activation neurons in the query vector and POI vector. The more of common indices of high activation neurons, the higher similarity between the query and POI.

\section{Conclusions}
\label{sec:conclusion}
\par
In this paper, we propose a novel deep learning model for POI retrieval. The model can effectively extract semantic features and compute the similarity score between the query and POI. The proposed DPSM model has three key properties: 
(1) DPSM makes full use of semantic information by using multi-granularity representation of texts; (2) DPSM uses an embedding layer before a convolution layer to speed up the computation; (3) DPSM uses the POI name and POI address to build models and obtain more complete information of POI. The experimental results show that our model achieves outstanding performance on the click dataset and the relevance dataset.

\bibliographystyle{ACM-Reference-Format}
\bibliography{sigiriti}

\end{document}